\documentclass[11pt]{article}
\usepackage{amsmath,amsfonts,amssymb}

\usepackage[total={6.5in,8.75in}, top=1.2in, left=0.9in, includefoot]{geometry}

\usepackage{graphicx}

\newcommand{\Sec}[1]{\S \ref{#1}}
\newcommand{\App}[1]{Appendix \ref{#1}}
\newcommand{\Fig}[1]{Fig.~\ref{#1}}
\newcommand{\Tbl}[1]{Table~\ref{#1}}
\newcommand{\ave}[1]{{\langle #1 \rangle}}

\newcommand{\Eq}[1]{(\ref{#1})}

\newcommand{\sn}{\text{sn}}
\newcommand{\cn}{\text{cn}}
\newcommand{\dn}{\text{dn}}

\newcommand{\sgn}{\text{sgn}}

\newcommand{\sD}{\mathcal{D}}

\newcommand{\sH}{\mathcal{H}}

\newcommand{\Z}{\mathbb Z}
\newcommand{\Bu}{\mathbf{u}}
\newcommand{\Bx}{\mathbf{x}}

\title{Chaotic Advection and the Emergence of Tori in the K\"uppers-Lortz State}
\author{Paul Mullowney, Keith Julien, and James~D.~Meiss \thanks
      {
         PM was supported by NSF VIGRE grant DMS-9810751.
         KJ was supported in part by NSF grant OCE-0137347.
         JDM was supported in part by NSF grant DMS-0707659.
      }
    \\
 \begin{tabular}{c}
	Department of Applied Mathematics\\
    University of Colorado \\
	Boulder, CO 80309-0526 \\
\end{tabular}
}

\begin{document}
\maketitle

\begin{abstract}
Motivated by the roll-switching behavior observed in rotating
Rayleigh-B\'enard convection, we define a K\"uppers-Lortz (K-L) state
as a volume-preserving flow with periodic roll switching.  For an
individual roll state, the Lagrangian particle trajectories are
periodic. In a system with roll-switching, the particles can exhibit
three-dimensional, chaotic motion. We study a simple phenomenological
map that models the Lagrangian dynamics in a K-L state. When the roll
axes differ by $120^{\circ}$ in the plane of rotation, we show that
the phase space is dominated by invariant tori if the ratio of
switching time to roll turnover time is small. When this parameter
approaches zero these tori limit onto the classical hexagonal
convection patterns, and, as it gets large, the dynamics becomes fully
chaotic and well-mixed. For intermediate values, there are interlinked
toroidal and poloidal structures separated by chaotic regions. We also
compute the exit time distributions and show that the unbounded
chaotic orbits are normally diffusive. Although the
map presumes instantaneous switching between roll states, we show that
the qualitative features of the flow persist when the model has
smooth, overlapping time-dependence for the roll amplitudes (the
Busse-Heikes model). 
\end{abstract} 

\section{Introduction}

\textbf{In this paper, we study the motion of a passive tracer in 
two simple kinematic models of the
K\"uppers-Lortz instability, a classical problem in rotating
convection. Each model is composed of sequentially acting rolls
whose axes lie in a plane and are rotated by $120^{\circ}$. In the limit
when the switching from roll to roll is rapid, the
tracer trajectories limit onto a classical hexagonal convection pattern. As
the switching time increases, tracer trajectories predominately lie on
two-dimensional tori. Eventually, these tori are destroyed and the dynamics
becomes chaotic and strongly mixing.}

The mixing of tracer elements in a flow is due to a combination of
stirring and diffusion.  While stirring rapidly transports the tracers
by kinematic advection, the associated stretching and folding of
material lines and surfaces ultimately triggers diffusive processes
that homogenize the tracers into a blended mixture.  Despite this
relatively simple description, a complete understanding of the spatial
and temporal evolution of tracer elements in a given flow remains 
a mathematically challenging problem \cite{Ottino1990}.  This is further
compounded if the tracers are not passively advected, but nonlinearly
interact with fluid flow and perhaps other species. The
phenomena of mixing is fundamentally important in many physical and
engineering applications and it occurs at a variety of scales
ranging from the very small (micrometer scale) to the very large (planetary
scales and beyond).  For instance, mixing in microchannels can be used to
efficiently homogenize reagents in chemical reactions even when the flow is
laminar \cite{Stroock2002}.   In reaction-diffusion and combustion systems, the
efficiency of mixing reactive agents is a crucial factor.  Understanding
transport for planetary scale flows is critical for climate modeling and
pollution dispersion in atmospheric science \cite{Brown1991} and eddy dynamics
in oceanography \cite{Jones2002}.  

While mixing is prominent in turbulent flows, effective mixing also occurs
in laminar flows that have chaotic Lagrangian trajectories. The
innovative work of Aref \cite{Aref1984}  gave birth to the subject of
{\sl chaotic advection}, which provides a mathematical foundation for
investigating mixing  and transport from a dynamical systems perspective
 \cite{Aref2002}. Aref called his seminal model the ``blinking vortex;"
it was specifically designed to yield nonintegrable Lagrangian
trajectories.  This model can be interpreted as an idealized mixing
protocol where passive tracers are successively captured by the
velocity fields of vortical stirrers that are the analogues of
turbulent eddies with finite lifetimes. 

Recently, several models for three-dimensional Lagrangian dynamics have been
constructed out of vortical roll arrays. Transport in a set of
orthogonal, time independent roll arrays was investigated in
\cite{Fogleman2001,Vargas2003}. In this system long-range,
superdiffusive transport only occurs along the 
direction perpendicular to the axes of both rolls. This flow could be
realized using a combination of thermal convection and
magnetohydrodynamic forcing.  
 
In a previous paper, we studied motion in a set of rolls whose axes
are orthogonal, and whose amplitudes oscillate
\cite{Mullowney2005}. This models the 
oscillations observed in experiments on convection in binary fluids
\cite{Moses1991}. The experiments showed that the convective
instability gives rise to a sequence of temporally alternating,
orthogonal roll arrays whose axes are parallel to the square
boundaries. The transition from one set of rolls to the 
other was observed to be rapid, so that much of the time the fluid is
dominated by a single set of parallel rolls. In our model, regardless
of the choice of temporal dependence of the roll amplitudes, the
Lagrangian dynamics is restricted two-dimensional surfaces;
consequently, our model does not lead to three-dimensional 
mixing, though the two-dimensional dynamics can be chaotic. Our model
contrasts with that of \cite{Fogleman2001,Vargas2003} in that their
rolls are $90^{\circ}$ out of phase.  

Roll-switching was predicted for rotating convection in thin fluids
by K\"uppers and Lortz \cite{Lortz1969,Kuppers1970}.  Here a
thin fluid layer of large horizontal extent (i.e. small aspect ratio)  
is rotated about its vertical axis.  When the rotation rate is small, 
a stationary roll pattern forms at a critical Rayleigh number. 
Above a critical rotation rate this pattern becomes unstable to a
second set of two-dimensional rolls whose 
axes are rotated with respect to the original roll axes by some
``switching angle," $\Theta$. These rolls grow, causing the first set
to decay. The secondary rolls then lose stability to a new set of
rotated rolls  and the process repeats.  

This behavior has been observed in a variety of experiments \cite{Hu1998}. 
Typically the weakly supercritical regime shows
large patches of homogeneous rolls that loose stability to rolls with
an axis rotated by a switching angle that varies with the Taylor and
Prandtl numbers. For example in water (Prandtl number near seven),
$\Theta=63\pm 3^{\circ}$ \cite{Zhong1991}. Theoretical investigations
confirm the dependence of $\Theta$ on rotation rate, boundary
conditions, and Prandtl number \cite{Clever1979,Clune1993}. Though the
time dependence of the roll switching is not observed to be precisely
periodic, there is a mean switching frequency that decreases as the Taylor  
number increases, but increases with the Rayleigh number \cite{Ning1993}.

Motivated by these experiments, we call three-dimensional fluid flow consisting of
a set of rolls with different axes, whose amplitudes oscillate a {\em
  K\"uppers-Lortz state}. In this paper we analyze a simple model of 
the Lagrangian trajectories in a K\"uppers-Lortz state. We begin in
\Sec{KLState} by constructing a time-dependent model with a simple
spatial roll structure. When the switching is much faster than the
roll turnover time, it can be idealized as instantaneous and the flow
can be viewed as a composition of volume-preserving maps corresponding to the action of
each individual roll. This gives rise to what we call a {\em blinking
  roll} model in \Sec{sec:mapping};  it is analogous to Aref's
blinking vortex model. To check the importance of the assumption of
instantaneous switching, we also introduce in \Sec{sec:bhmodel} a
continuous model using the Busse-Heikes system of odes for the
evolution of the roll amplitudes \cite{Busse1980,Toral2000}.  When the
switching angle $\Theta$ is $120^\circ$ or $60^\circ$ symmetry, as
reviewed in \Sec{sec:Geometry}, allows us to restrict our analysis to
a fundamental hexagonal cell. Finally, in \Sec{sec:Numerical},
computational results are discussed and mixing is analyzed using Lyapunov
exponents and anomalous diffusion exponents \cite{Klafter1996}.

Even when some three-dimensional mixing occurs in these models, it is
usually not complete because there are families of two-dimensional
invariant tori that form barriers to the Lagrangian trajectories. Such
families of tori are prominent in three-dimensional, incompressible
flows; they play a roll analogous to that of invariant closed curves in
two-dimensional, area-preserving mappings \cite{Meiss1992}. Tori have
also been experimentally observed, for example, in a cylindrical
container with a tilted stirrer
\cite{Fountain1998,Fountain2000}. These are visualized  with sheet
lasers and fluorescent dyes and correlate closely with corresponding
island chains of the Poincar\'e sections of a model flow. Similar tori
have also been seen for weakly buoyant tracers in a laminar flow
\cite{Shinbrot2001}. We investigate some of the behavior of the
families of tori in our system in \Sec{sec:Numerical}. 


\section{K\"uppers-Lortz Roll Model}\label{KLState}

For the purpose of making progress analytically, we use the simplest roll model
for rotating convection: the linearized solutions in the Boussinesq
approximation with stress-free boundary conditions \cite{Chandrasekhar1961}. 
These rolls are tilted by an angle $\eta$ due to the Coriolis force. In scaled coordinates with the fluid confined to $z \in [\pi/2, \pi/2]$, the roll velocity field is
\begin{eqnarray}\label{eq:TRoll}
        \Bu_\eta &=&A \left(-\tan(\eta)\sin(y)\sin(z), -\sin(y)\sin(z),
        -\cos(y)\cos(z)\right)\;. 
\end{eqnarray}
Here, $A$ is the amplitude of the rolls and 
\begin{equation}\label{eq:tilt}
        \tan(\eta) =\frac{\tau}{k_c^2+\pi^2} \;,
\end{equation}
where $\tau$ denotes the Taylor number and $k_c$ is the critical wave
number at the onset of convection. For highly viscous flows and
infinite Prandtl number, roll switching starts at a Taylor number
$\tau_c=54.8$ with  $k_c=3.95$, so that
$\eta\approx65^{\circ}$ \footnote  
{
  In this case, the switching angle for the K\"uppers-Lortz state is $\Theta=59.7^{\circ}$ \cite{Lortz1969}.
}.
When the thin fluid layer is
not rotating, $\tau=\eta=0$, and the velocity field simplifies to:
\begin{equation}\label{eq:Roll}
        \Bu_0= A \left(0,-\sin(y)\sin(z), -\cos(y)\cos(z)\right)
\end{equation}
Although a rigid lower boundary together with a free upper surface is more
physical (in which case Chandrasekhar functions would be the
appropriate building block \cite{Chandrasekhar1961}), this choice would complicate both the analytical
and numerical treatment. Moreover, in the limit $\tau\gg
1$, it has been shown \cite{Julien1998,Niiler1965} that the
stress-free solutions converge to the rigid ones outside the passive 
$\mathcal{O}(\tau^{-\frac{1}{2}})$ Ekman boundary layers. 

Trajectories of \Eq{eq:TRoll} are confined to
two-dimensional contours about axes parallel to $\hat{e}_1$ (the
$x$-axis). 
A K\"uppers-Lortz state can be modeled in a simple way 
by using the roll solutions \Eq{eq:TRoll} as the building blocks of
the flow. Our model is a superposition of rolls with amplitudes $A_j(t)$ whose axes are rotated by angles that are multiples of $\Theta$:
\begin{equation}\label{eq:velField}
  \Bu(\Bx,t)=\sum_{j=0}^{N-1}A_{j+1}(t)R\left(j\Theta\right)\Bu_\eta\left(R^T(j\Theta)\Bx\right)\;.
\end{equation}
Here $R$ is the rotation about the $z$-axis by an angle
$\Theta$: 
\begin{equation}\label{eq:defineR}
        R(\Theta)=
        \left(\begin{matrix}
         \cos(\Theta) & -\sin(\Theta) & 0 \cr 
         \sin(\Theta) & \cos(\Theta) & 0 \cr
         0 & 0 & 1
        \end{matrix}\right)\;.
\end{equation}
Since each individual roll is incompressible, any composition with arbitrary time-dependence is also incompressible. 

We will assume that the rolls act sequentially. That is, when the $j^{th}$ roll is active, the amplitudes of all other rolls are nearly zero, and that as $t$ increases, $A_j(t)$ decreases to zero, and the next roll $A_{j+1}(t)$ activates and so forth.

If $\frac{\Theta}{2\pi} = \frac{p}{q}$ is rational, then it is appropriate to set $N =q$, since if the $j^{th}$ roll looses stability to the $(j+1)^{st}$ roll, then the whole process starts over after $q$ steps. If, however, $\frac{\Theta}{2\pi}$ were irrational, then the instability would continue to generate rolls at angles $j\Theta$, and the the ``number" of rolls, $N$ would be effectively infinite.

The simplest case of \Eq{eq:velField} corresponds to $N=2$ and to orthogonal rolls, as we studied in \cite{Mullowney2005}. Here we will study the case $N=3$ and a switching angle $\Theta = 2\pi/3$. Letting
\[
	R \equiv R(2\pi/3)
\]
we then have the velocity field
\begin{equation}\label{eq:VelField}
  \Bu_T(\Bx,t)=A_1(t)\Bu_\eta(\Bx)+A_2(t)R\Bu_\eta(R^{T}\Bx)+
  		A_3(t)R^{T}\Bu_\eta(R\Bx) \;;
\end{equation}
Note that this model corresponds to a switching angle of $120^{\circ}$, 
not $60^{\circ}$. However, we shall see in \Sec{sec:Geometry} that the
roll \Eq{eq:TRoll} has a reflection symmetry through the origin in 
the $xy$ plane, thus these two are equivalent.  

When the amplitudes are constant and equal, this velocity field gives
the (rotating) classical hexagonal convection cell
\begin{equation}\label{eq:HexVelField}
        \Bu_{h}(\Bx)=\Bu_\eta(\Bx)+R\Bu_\eta(R^T\Bx)+R^T\Bu_\eta(R\Bx)\;,
\end{equation}
The orbits of this autonomous system are either periodic or heteroclinic connecting 
equilibria on the upper and lower boundaries \cite{Chandrasekhar1961}.


\section{The Blinking Roll Map}\label{sec:mapping}

The simplest model of sequential rolls corresponds to roll amplitudes that are periodic step functions. Motivated by
the experiments, as well as by simplicity, we assume that both the strength and the  the activation time of each roll is the same. In this case we can rescale time, setting $At \to t$ so that the roll amplitudes become one. The scaled roll activation time, $T$, thus represents both the strength and period of the flow. In this case the roll amplitudes are
\begin{equation}\label{eq:defineA}
  A_{j}(t)=\left\{\begin{array}{ll} 1, &\mbox{ for } (j-1)T \le t < jT \;,\cr
    					0, & \mbox{ otherwise}
  \end{array}\right.\;.
\end{equation}
for $0 \le t < NT$ and periodically repeating thereafter.

When the amplitude is constant, the flow for the vector
field \Eq{eq:Roll} can be solved exactly in terms of Jacobi elliptic
functions. To get the flow for the tilted roll, we use the $\eta = 0$ solutions of \cite{Mullowney2005} and note note that in \Eq{eq:TRoll},
$\dot{x}=\tan(\eta)\dot{y}$, giving
\begin{equation} \label{eq:TiltMap}
         \Phi_{t}(\Bx) =
         \left( \begin{array}{c}
                x +\tan(\eta)\left[\sgn(y)\cos^{-1} \delta -y\right]\\
                \sgn(y)\cos^{-1}\delta\\
                \sin^{-1}\chi
        \end{array} \right) \;,
\end{equation}
where
\begin{align*}
        \delta(y,z,t) & \equiv \frac{\cos(y)\cn(t)\dn(t)+\sn(t)\sin(z)\sin^2(y)}
                {1-\cos^2(y)\sn^2(t)} \;, \\
        \chi(y,z,t) & \equiv \frac{\sin(z)\cn(t)\dn(t)-\sn(t)\cos(y)\cos^2(z)}
                {1-\sin^2(z)\sn^2(t)} \;, 
\end{align*}
and the elliptic functions have modulus
\begin{equation}\label{eq:modulus}
     k = \sqrt{1-\sin^2 y \cos^2 z} \;.
\end{equation}
By setting $t=T$ in \Eq{eq:TiltMap}, we get the time-$T$ map 
\begin{equation}\label{eq:Fdefine}
 F(\Bx) = \Phi_{T}(\Bx)\;.
\end{equation}
The effective map for the other rolls can be obtained from
\Eq{eq:Fdefine} by rotation.  For example in the three-roll case the resulting mapping is
\begin{align}\label{eq:HexMap}
 f&=(R\circ F)^3 \;.
\end{align}
The map $f$ corresponds to the flow of \Eq{eq:VelField} over the time $3T$. Note that since the vector field is incompressible, the map $f$ is volume preserving. Thus, the determinant of the Jacobian, $Df$, is equal to one.

Different values of $\Theta$ can can be easily accommodated by the map by replacing
$R$ with $R(\Theta)$. When $\frac{\Theta}{2\pi}$ is irrational, we simply repeatedly iterate $R(\Theta) \circ F$.


\section{Continuous-Time Models}\label{sec:bhmodel}

One simple model of the temporal dependence of the amplitudes of three rolls is
the Busse and Heikes model \cite{Busse1980},
\begin{align}\label{eq:BH}
 \dot{A}_1&=A_1(1-|A_1|^2-\alpha |A_2|^2-\beta |A_3|^2) \nonumber \\
 \dot{A}_2&=A_2(1-|A_2|^2-\alpha |A_3|^2-\beta |A_1|^2) \\
 \dot{A}_3&=A_3(1-|A_3|^2-\alpha |A_1|^2-\beta |A_2|^2) \nonumber \;.
\end{align}
When $\alpha$ and $\beta$ are positive, this system has a heteroclinic cycle 
between three states. This model could also include evolution of the phases of the rolls, but we assume that these are constant. 

A thorough review of the dynamic properties of \Eq{eq:BH} is given in
\cite{Toral2000}. When $\alpha > 1 > \beta$, there exists
an attracting heteroclinic cycle connecting the saddle equilibria,
i.e. the pure roll states: $(A_1,A_2,A_3)=(1,0,0)$, $(0,1,0)$, and
$(0,0,1)$. The presence of the attracting equilibria yields a system
that spends increasing amounts of time around the unstable roll states. 
This feature is not physical, since experimental observations suggest a
characteristic period of oscillation associated with a 
K\"uppers-Lortz state. Busse and Heikes offered several ideas for overcoming this problem.  
One proposal involves the introduction of small amplitude noise into \Eq{eq:BH}
designed to simulate the missing terms in the true amplitude equations.
Noisy perturbations of attracting heteroclinic cycles were first studied
in \cite{Stone1990}. Perturbations of this type prevent the system from
spending increasing amounts of time about any one of the saddle equilibria \cite{Toral2000}. 
The noisy system has a statistically ``stable'' limit cycle with a well-defined 
mean period. A second suggestion is to add slow spatial variation to the amplitude 
\cite{Tu1992}. Adding terms of form
$\partial^2_{x_i}A_i$ to each equation in \Eq{eq:BH} respectively also creates a 
stable oscillation, although its period is difficult to predict.

We implement the stochastic method by adding the derivative
of small white noise terms, $\xi_i(t)$, to \Eq{eq:BH}.
Specifically, the $\xi_i(t)$ are Wiener processes with mean
$2\epsilon$ and variance $\epsilon^2$. The mean is nonzero to help avoid
pushing the amplitudes out of the positive octant \footnote
{
  Though there is a small chance of getting kicked into the unphysical quadrants when
  the integrations are close to the coordinate planes,  in
  practice, choosing mean $2\epsilon$ appears sufficient to prevent this.
}.
Time series are obtained using a second-order stochastic Runge-Kutta solver
\cite{Kloeden1992,Burrage1996}. Noise causes the orbit to be pushed away from the
unstable single roll solutions, preventing it from spending
arbitrarily long durations near any of those states. 

\begin{figure*}
\begin{center}
\includegraphics[width=6in]{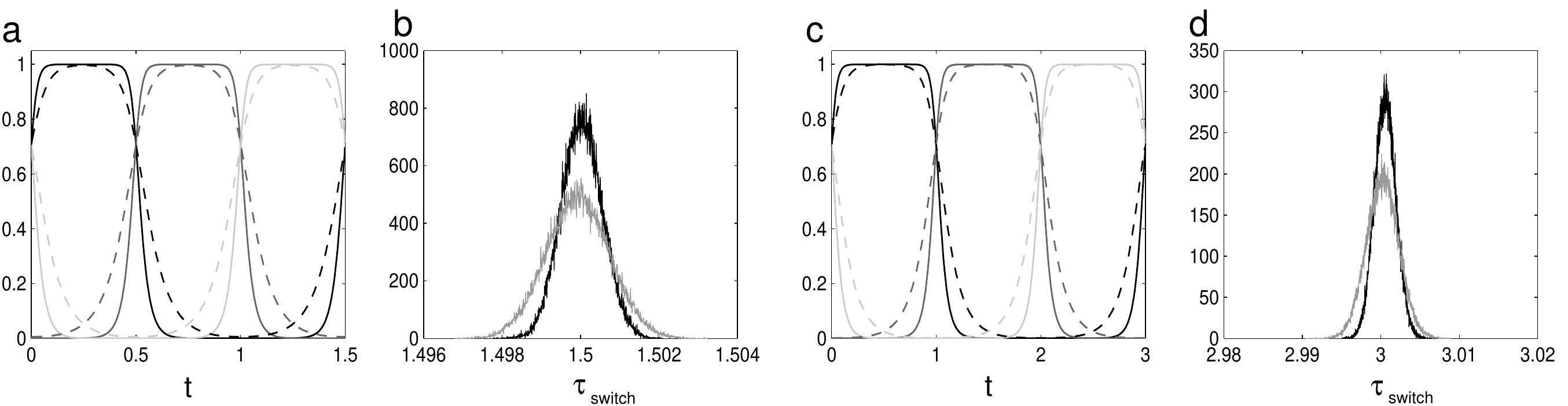}
\end{center}
\caption{Stochastic Busse-Heikes model with parameters given
in \Tbl{Table1}.
  (a) Evolution of the amplitudes: the solid lines correspond to
  (case a) in \Tbl{Table1} and the dashed lines to (case b). 
  (b) Period PDF for an integration time  $t_f=1.5(10)^5$.  
  The black curve (case a) has $\tau_{\text{switch}}=1.500$,
  standard deviation $\sigma=.0005$, skewness $\zeta=-.004$, and kurtosis $\kappa=-.022$. 
  The gray curve (case b) has $\tau_{\text{switch}}=1.500$, $\sigma=.0008$,
  $\zeta=-.017$, and $\kappa=.057$. 
  (c) Amplitudes for, (case e), solid lines, and (case f), dashed lines. 
  (d) Period PDF for $t_f=3(10)^5$.   
  For (case e), black, $\tau_{\text{switch}}=3.00$,
  $\sigma=.001$, $\zeta=.057$, and $\kappa=.002$, while for (case f), gray, 
  $\tau_{\text{switch}}=3.000$, $\sigma=.003$, $\zeta=-.006$, and $\kappa=.018$.} 
\label{Figure4}
\end{figure*}

Detailed pictures of the asymptotically periodic evolution are 
shown in \Fig{Figure4}. The model has three parameters $\alpha$, 
$\beta$, and $\epsilon$ that can be chosen to control mean period, shape, and symmetry
of the amplitude waveforms. 
The curves in \Fig{Figure4}a and c show typical realizations of the amplitude 
evolution; the solid curves correspond to much smaller values of
$\epsilon$ than the dashed curves \footnote
{
  In the numerical work of \Sec{sec:Numerical},  
  the waveforms in \Fig{Figure4}a and c will be used with tilt angles
  $\eta=65^{\circ}$ and $\eta=0^{\circ}$ in \Eq{eq:VelField} respectively.
}. 
In \Fig{Figure4}b and d, probability distribution functions (PDFs)
of the periods are shown.  For the first case the mean period, $\tau_{\text{switch}}$
is close to $1.5$, while in the second it is close to $3$. For each
case, a step size, $\Delta t$, was chosen so that there were roughly
$300$ steps per period. The PDFs were computed  
using $\sim10^5$ periods of a single realization. In both cases,
the probability distribution functions for the period appear to be nearly Gaussian.

Let $\tau_{\text{roll}}$ denote the typical scaled turnover time
of a single roll (when $A=1$, $\tau_{\text{roll}} =\pi^2$, 
see Appendix \ref{sec:meanT}).
The time $\tau_{\text{switch}}$ measures the effective strength of 
the rolls; consequently when
\begin{equation}\label{eq:limit1}
 \frac{\tau_{\text{switch}}}{\tau_{\text{roll}}}<<1\;,
\end{equation}
the strength of the rolls is relatively weak. As this parameter grows 
the rolls become stronger. 

Let $\tau_{\text{trans}}$ denote
the typical transition time between roll states (e.g., the transition
time between $0^{\circ}$ and $120^{\circ}$ states). 
If the ratio of transition time to switching time is small:
\begin{equation}\label{eq:limit2}
 \frac{\tau_{\text{trans}}}{\tau_{\text{switch}}}<<1\;,
\end{equation}
then a relatively large time is spent in a
particular roll state compared to the transition to the next roll. 
The parameter $\epsilon$ is the key to controlling this ratio; indeed,
comparing the solid and dashed curves in in \Fig{Figure4}, it is clear 
that as $\epsilon$ grows this inequality becomes invalid.


\section{Geometry}\label{sec:Geometry}

Since the vertical component of the velocity \Eq{eq:TRoll} vanishes at $z = \pm \pi/2$, the orbits 
are confined to the infinite domain
\begin{equation}\label{eq:domain}
  \sD=\left\{(x,y,z):-\frac{\pi}{2}\leq
    z\leq\frac{\pi}{2}\right\} \;.
\end{equation}

\begin{figure}
\begin{center}
\includegraphics[width=4in]{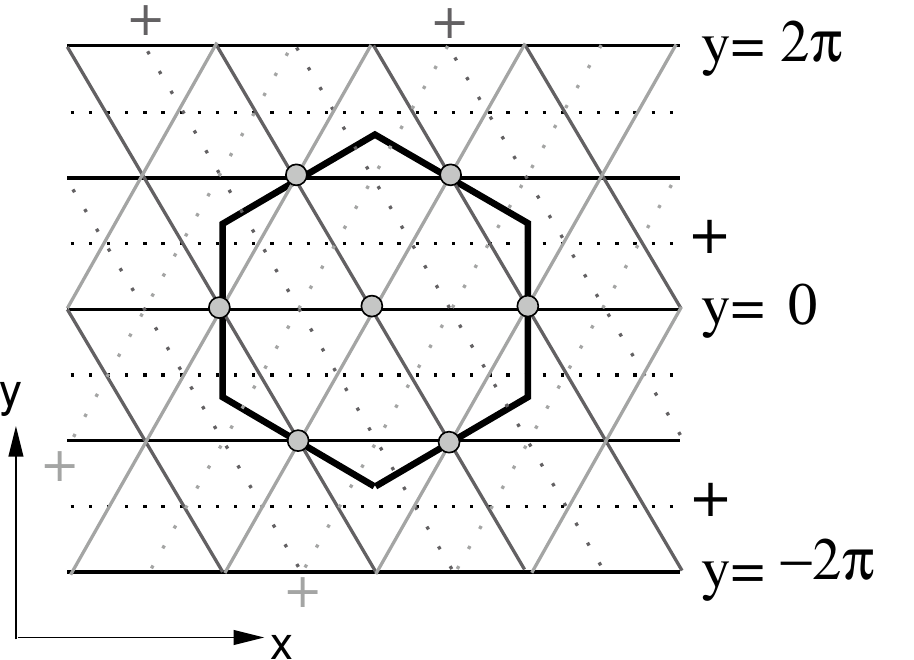}
\end{center}
\caption{Schematic of the roll geometry in $z=0$ plane. The open circles
  correspond to saddle equilibria at $z=\pm\frac{\pi}{2}$. The solid lines
  correspond to roll boundaries and the dotted lines correspond to roll
  axes. The $+$ signs denote right-handed or positive rotation about its
  axis. The hexagon, $\sH$ is a fundamental cell of the flow.} 
\label{Figure5}
\end{figure}

As we will review below, the velocity field \Eq{eq:VelField} is invariant under a discrete
translational symmetry generated by the unit vectors
\[
        \hat{u}_1=\left(1,0,0\right) \;,\quad \hat{u}_2=\left(\frac12,\frac{\sqrt{3}}{2},0\right) \;.
\]
These vectors generate the planar ``wallpaper" group denoted by $p3$:
\begin{equation}\label{eq:p3Group}
        p3 = \left\{t_{mn}(\Bx) = \Bx + \frac{4\pi}{\sqrt3}( m\hat{u}_1+n \hat{u}_2) :
                         m,n \in \Z \right\} \;.
\end{equation}
corresponding to a single, order-three axis of symmetry. Here we have
scaled the length of the vectors to correspond with the unit cell size
in our examples. This group has a fundamental cell that is a regular
hexagon, $\sH$, of width $4\pi/\sqrt3$ and height $8\pi/3$, see
\Fig{Figure5}.

Regardless of the choice of time dependence for the amplitudes, the vector 
field \Eq{eq:VelField} has a fourteen saddle equilibria on the upper and 
lower boundaries, $z = \pm\frac{\pi}{2}$; these are shown as the dots in 
\Fig{Figure5}. Each vertically separated pair is connected by an invariant 
line giving a heteroclinic connection. For example, the saddles
at $\left(0,0,\pm\frac{\pi}{2}\right)$ are connected
by the invariant line $\{x=y=0\}$.  Along this heteroclinic connection, 
the fluid motion is downward. The remaining twelve saddles can be found on the boundary of
$\sH$, six at $z=\pi/2$ and six at $z=-\pi/2$. There are heteroclinic orbits 
connecting these pairs that move upward. For
example, there is an upward moving heteroclinic connection between the points
$\left(\frac{2\pi}{\sqrt{3}},0,-\frac{\pi}{2}\right)$ to
$\left(\frac{2\pi}{\sqrt{3}},0,\frac{\pi}{2}\right)$.

Note that since the map \Eq{eq:HexMap} is an exact solution of 
\Eq{eq:VelField} for the special case of step function amplitudes, 
the equilibria and heteroclinic connections of the flow become fixed 
points and heteroclinic lines of the map.  


\section{Numerical Explorations}\label{sec:Numerical}

In this section we study the mixing and transport for the flow
\Eq{eq:VelField} and compare it to that for the blinking roll mapping
\Eq{eq:HexMap}. One local measure of mixing corresponds to the Lyapunov
exponents as a function of initial condition. There is a correlation
between large exponents and regions of strong mixing. On the other hand
in regions  with many invariant tori, the exponents are zero or near zero.

A  second measure is the rate of particle drift throughout the domain. A
first goal is to distinguish between bounded and unbounded orbits. 
We will study those bounded orbits that
remain within the fundamental domain $\sH$---we call
these orbits ``trapped." Generally the ``regular" orbits appear to be
trapped (for example, those on invariant tori), especially for the case
$T_1=T_2=T_3$; however, there appear to be bounded invariant structures
larger than a single fundamental domain when the mapping times are not
equal. We will also compute the exit time distribution from a
fundamental domain. Transport of unbounded orbits can be quantified by
computing the growth rate of distance with $t$ and comparing this to
diffusive behavior. Since the trajectories are bounded in $z$, we
measure the drift in $(x,y)$ by looking at the growth rate of $r^2 = x^2
+ y^2$.

\subsection{Lyapunov Phase Portraits}\label{sec:Lyapunov}

In this section, we compute the three Lyapunov exponents, $(\mu_1,\mu_2, \mu_3)$, 
as a function of initial condition for both the flow and map models. 
Ideally we would 
compute the exponents for each initial condition on a three-dimensional 
mesh covering the fundamental cell, but this would be
prohibitively expensive in computation time and the results would be
difficult to display. Instead, we use two, two-dimensional slices in the
fundamental cell $\sH$: the horizontal midplane, $z=0$, and the vertical
cross section, $y=0$. The horizontal slice is hexagonal and is divided
into a $70\times70$ grid and the vertical slice is rectangular and
divided into a $100\times50$ grid. 

\begin{figure*}
\begin{center}
\includegraphics[width=5.5in]{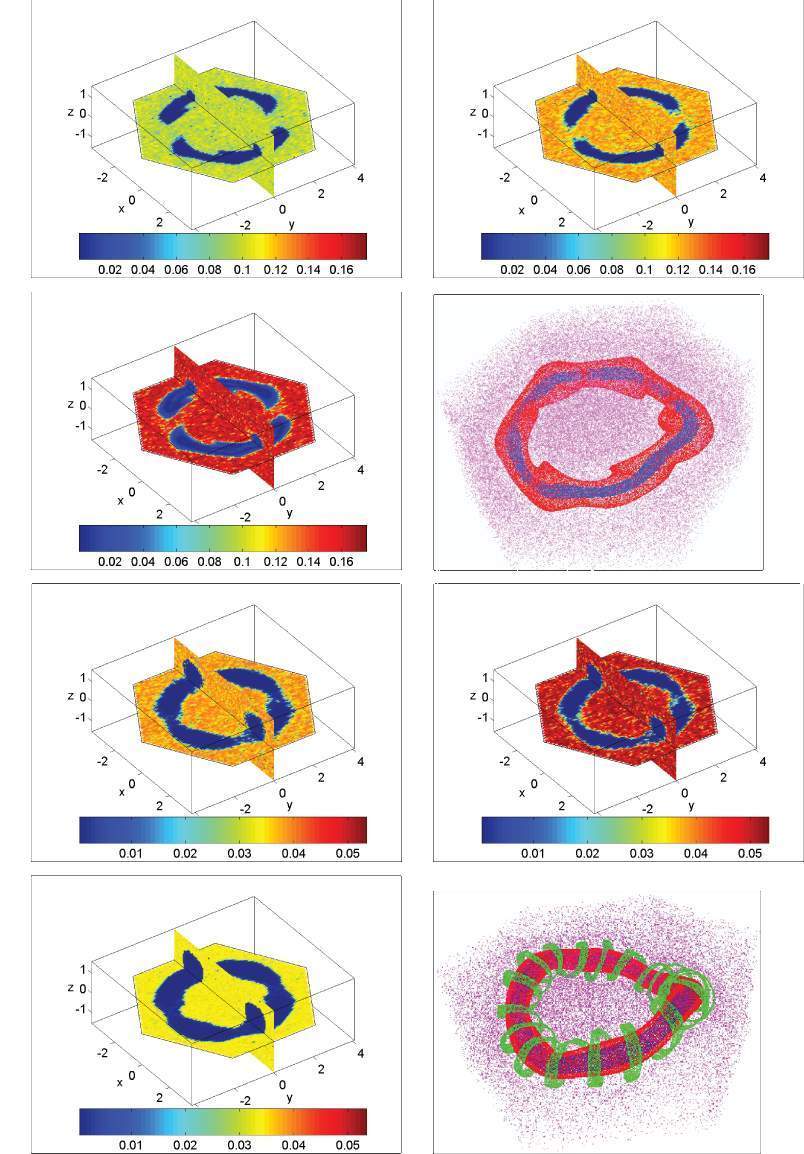}
\end{center}
\caption{Lyapunov phase portraits for the flow \Eq{eq:VelField} and map
  \Eq{eq:HexMap} for the parameter values given in
  \Tbl{Table1}. The largest Lyapunov exponent is computed by
  integrating or iterating for $10^5$ periods and the graphed values
  are obtained by averaging over the final $100$ periods. The
  color scale for $\mu_1$ is indicated at the bottom of each pane. Sample
  trajectories of the mapping are given in panels (d) and (h) showing
  regions occupied by invariant tori. Panels (a),(b), and (c) have the
  same color scale as do panels (e), (f), and (g).} 
\label{Figure6}
\end{figure*}

Since the map is volume-preserving and the flow is incompressible, the
sum $\mu_1+\mu_2+\mu_3=0$ for both models. 
In \Fig{Figure6}, we show only the value of the largest exponent $\mu_1$ 
for each initial condition on these grids. We call these pictures ``Lyapunov phase
portraits;" they help to differentiate between chaotic and regular
regions and provide a local measure of chaos that correlates with rapid
mixing. The exponents are computed using the QR method of Rangarajan et
al \cite{Rangarajan1998,Habib1999}. 
The second and third Lyapunov exponents were also computed for all cases.
Typically we observed that $|\mu_2| \sim10^{-4}$ which is two orders of magnitude smaller than $\mu_1$ in the chaotic region. Given that only $10^5$ iterations were used in the computation and that converge is slow, a longer integration might be expected to yield an evan smaller second exponent. Thus our results are consistent with $\mu_2=0$, and consequently with $\mu_3=-\mu_1$. 
Note that an autonomous flow always has a zero exponent. More generally, if the flow or
map were reversible then we would also expect $\mu_2 = 0$. We have not been able to 
show, however, that our systems are reversible, and 
thus do not have an explanation for the near vanishing of $\mu_2$.

By computing the Lyapunov phase portraits for both the flow
\Eq{eq:VelField} and the map \Eq{eq:HexMap}, a direct comparison can be
made between the two dynamical systems if the parameter values are
equivalent. For the flow we use a full integration of the stochastic
Busse-Heikes model to generate the amplitude waveforms; those shown in
\Fig{Figure4} give a typical cycle. Since there are three blinking-rolls per map iteration,
the corresponding parameter $T$ for the mapping is one third of the
average period of the amplitude functions.

The results are shown in
\Fig{Figure6} and the corresponding parameter values are given in
\Tbl{Table1}. In each case the total time corresponds to approximately 
$10^5$ oscillation cycles. The average period of oscillation is 
approximately $1.5$ for the
first two flow cases (panels (a) and (b)), and we choose the
corresponding mapping time in panel (c) to be $0.5$ so that the total
integration time represented by iterating the map $10^5$ times is the
same as that for the flow. For panels (e) and (f) the average period is
$3.0$, and  the corresponding mapping time $T = 1.0$ is used in panel
(g).
\footnote{The average computation time was about $4$ cpu days for the
  flow. Computations were performed on a 4 processor, 64 bit,
  Itanium 2 1.3GHz (32KB L1 cache, 256 KB L2 cache, 3MB L3 cache)
  architecture. In contrast, the Lyapunov exponents for the discrete
  maps took $2-3$ cpu hours on the same machine.}

Results for the flow with $\eta=65^\circ$ are displayed in
\Fig{Figure6}a and \Fig{Figure6}b. The amplitude waveforms in
these two cases differ in their shapes, as shown by the solid and dashed
curves in \Fig{Figure4}. The images indicate that there are toroidal
regions of near zero Lyapunov exponent surrounded by a chaotic sea. The
breaks in the toroidal region on the hexagonal slice correspond to
locations where the torus does not intersect this slice. It is striking
that the tori exist even when the three amplitude waveforms have
significant intervals of overlap as they do for \Fig{Figure6}b. This
suggests that the details of the amplitude wave form is not a critical
factor in the existence of the tori. The corresponding Lyapunov phase
portrait for the map, shown in \Fig{Figure6}c, is visually quite
similar to  (and as shown in \Tbl{Table2} has a high correlation with)
the first two flow cases. In \Fig{Figure6}e and 
\Fig{Figure6}f, the Lyapunov phase portraits are shown for the
waveforms in \Fig{Figure4}c with $\eta=0^{\circ}$. Once again, these
portraits show a high correlation with those for the corresponding
mapping, shown in \Fig{Figure6}g. Although the $\eta=0^{\circ}$ case
is nonphysical for rotating convection, it is interesting to see that
the toroidal structures persist for small $\eta$ and are robust as
this parameter varies.  

\begin{table}
\caption{Parameters for the Lyapunov phase portraits of \Fig{Figure6}} 
\begin{center}
\begin{tabular}{|c|c|c|c|c|c|c|c|c|c|c|c|}\hline
\Fig{Figure6} & Model &$\eta$ & $\alpha$ & $\beta$ & $\epsilon$ & $\Delta t$ & Period & $T$ \\ \hline
a & flow & $65$ & $29.9205$ & $-27.9005$ & $10^{-14}$ & $.005$ & $1.5$ & \\ \hline
b & flow & $65$ & $12.7503$ & $-10.7303$ & $10^{-7}$ & $.005$ & $1.5$ & \\ \hline
c,d & map & $65$ & & & & & & $.5$ \\ \hline
e & flow & $0$ & $18.0225$ & $-16.0025$ & $10^{-16}$ & $.01$ & $3$ & \\ \hline
f & flow & $0$ & $5.7560$ & $-3.7360$ & $10^{-8}$ & $.01$ & $3$ & \\ \hline
g,h & map & $0$ & & & & & & $1$ \\ \hline
\end{tabular}
\end{center}
\label{Table1}
\end{table}

A few orbits inside the regions of low chaos (red, blue, and green) and
one in the chaotic region (purple) are shown in \Fig{Figure6}d for
\Eq{eq:HexMap}. The regions of near-zero
Lyapunov exponent correspond to a set of nested two-dimensional tori. At
the center of this nested set is apparently an elliptic, invariant
curve. The green orbit in \Fig{Figure6}h appears to correspond to a
torus that encloses an invariant curve that is helically wound about
the central family of tori.

In each case, the close correlation between the Lyapunov exponents of
the flow and mapping is clear from the pictures. To quantify this
connection, we computed the correlation coefficient between the values
of $\mu_1$ for mapping and for the flow on each slice. The results are
given in \Tbl{Table2}; coefficients near one indicate similar
structure whereas coefficients near zero suggest large differences between the
data sets. In each case, a reasonably strong correlation is obtained. It
is notable that as the ratio \Eq{eq:limit2} increases, 
the correlation drops \footnote
{
        Here $\tau_{trans}$ is
        computed by finding the total time $t$ that
        $A_i(t)>.95$ or $A_i(t)<.05$ for $i=1,2,3$. This
        value is subtracted from $\tau_{\text{switch}}$.
}.  
This seems to predominantly occur near
the outer boundary of the regular regions where the step function
approximation to the amplitudes does a poor job of approximating the
flow: the map tends to have a larger region foliated by invariant tori
than does the flow.

\begin{table}
\caption{Correlation Coefficient of largest Lyapunov exponent between the map and the flow}
\begin{center}
\begin{tabular}{|c|c|c|c|}\hline
        Comparison & $\tau_{trans}/\tau_{switch}$ & $y=0$ slice & $z=0$ slice \\ \hline
        (a) vs. (c) & .19 & .70 & .88 \\ \hline
        (b) vs. (c) & .47 & .69 & .86 \\ \hline
        (e) vs. (g) & .16 & .86 & .94 \\ \hline
        (f)  vs. (g) & .38 & .74 & .88 \\ \hline
\end{tabular}
\end{center}
\label{Table2}
\end{table}

\begin{figure*}
\begin{center}
\includegraphics[width=6in]{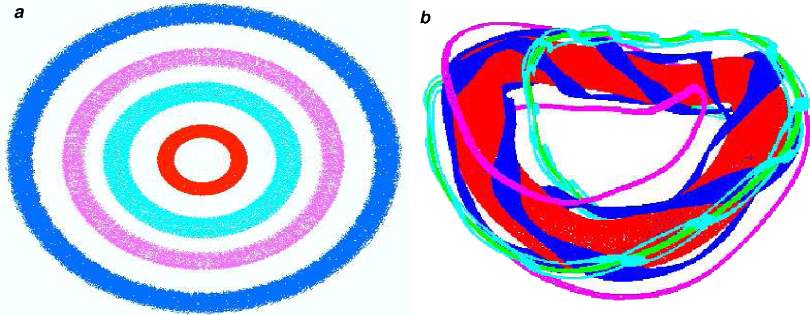}
\end{center}
\caption{Phase portraits of \Eq{eq:HexMap} for
  (a) $T=.2$, $\eta=65^{\circ}$, and $\Theta=122^{\circ}$, and (b)  $T=1$,
  $\eta=0^{\circ}$, $\Theta=45^{\circ}$. We plot points of
  the map $RF$ for each iterate instead of every third iterate as in \Eq{eq:HexMap}.} 
\label{Figure7}
\end{figure*}

As computations with the map are much quicker than those for the flow it
is easier to investigate the effect of changing parameters on the
map. Moreover, changing the switching angle $\Theta$ from  $120^{\circ}$
is problematic for the flow since the three-mode model \Eq{eq:BH} no
longer applies. It is very easy, however, to change the switching angle
in the map by choosing the rotation angle for the transformation $R$ in
\Eq{eq:HexMap}.  One of the immediate observations is that tori are
robust features of the mapping as the switching and tilt angles and time
step $T$ are varied. For example, a large family of tori can be seen
\Fig{Figure7}a (viewed from a camera looking down from large $z$). For
this case the ratio \Eq{eq:limit1} is small. The  size of
the innermost torus shown is on the order of one hexagonal convection
cell; however, since $\Theta\neq120^\circ$ the system no longer has 
hexagonal symmetry. Outside these tori, the motion is chaotic and diffusive (not
shown in the figure). 

A final example in \Fig{Figure7}b shows a number of helical tori
enclosing the central family. In the center of the light gray
orbits are invariant curves that wind twice in the toroidal direction
before returning to their original positions. It is possible that these
families of tori are created by period-doubling bifurcations of the
invariant curve at the center of the main torus or a resonant
bifurcation of a two-torus, but we have not verified this.

There are too many parameters in the system to completely explore the
full range of possible dynamics. A critical point is that the discrete
symmetry does not produce the tori: these structures are features of the
flow for many parameter values. 

\subsection{Breakdown of Integrability}

When the time $T$ is very small, it is easy to see that the map
\Eq{eq:HexMap} is, to first order in $T$, the same as the time-$T$ map
of the autonomous field \Eq{eq:HexVelField}. Thus for a fixed total
iteration time, we expect that the map dynamics will limit on the
integrable flow of \Eq{eq:HexVelField} as $T \rightarrow 0$. Indeed, we
observe that the mapping shows the same closed, integrable trajectories
of the classical hexagonal convection patterns
\cite{Chandrasekhar1961} when $T$ approaches zero for a given fixed   
total time $nT$. 

\begin{figure*}
\begin{center}
\includegraphics[width=6in]{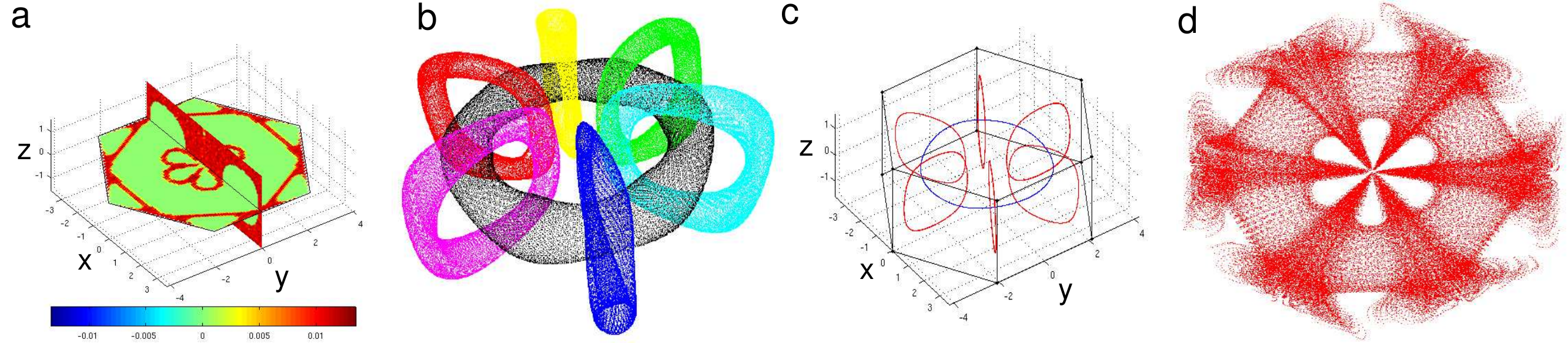}
\end{center}
\caption{Phase portraits of \Eq{eq:HexMap} for $\Theta = 120^\circ$, $\eta = 0$,
  and $T = 0.2$. Panel (a) shows the Lyapunov phase portrait,
  (b) the elliptic invariant curves at the center of the poloidal tori
  and (c) a chaotic trajectory  in the region between the tori.} 
\label{Figure8}
\end{figure*}

As $T$ grows, the closed trajectories are destroyed and are replaced by
nested invariant tori separated by small zones of chaos as illustrated
in \Fig{Figure8}. This example shows six families of ``poloidal"
two-tori linked with a central ``toroidal'' family. The poloidal tori
can be seen crossing the $z=0$ section in the 
Lyapunov phase portrait of panel (a) near the center and edges of the
hexagon. Invariant curves at the center of the poloidal and toroidal
families of tori, shown in panel (b), can be found using the simple
algorithm discussed in \App{sec:algorithm}. A trajectory in the chaotic
region between the tori is shown in a top down view in panel (c). This
domain can also be seen as the light grey regions in the Lyapunov phase
portrait of panel (a). 

\begin{figure*}
\begin{center}
\includegraphics[width=6in]{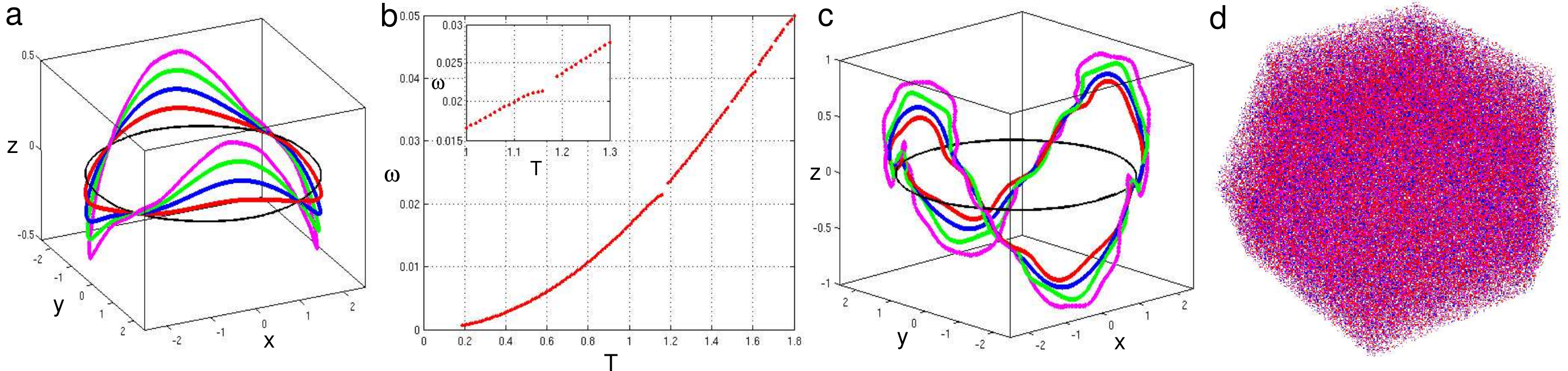}
\end{center}
\caption{Toroidal invariant curves of \Eq{eq:HexMap} for $\eta = 0$.
  (a) Curves for $T=0.2, 0.6, 1, 1.4$, and $1.8$. The solid black line is
  \Eq{eq:Equilibria}. (b) Rotation Number $\omega$ as a
  function of mapping time $T$. (c) Phase portrait for $T=2$ where there
  are no invariant tori giving complete mixing.} 
\label{Figure9}
\end{figure*}

This same structure is seen when $T$ is small for any value of $\eta$.
As $\eta$ increases the tori becomes increasingly twisted in the same
manner as the classical hexagons. For each $\eta$, as $T$ is further
increased, the six poloidal tori disappear leaving only the toroidal
family. Eventually the toroidal family is destroyed as well and the cell
appears to be completely chaotic (e.g. \Fig{Figure9}d below). In
\Tbl{tbl:InvTori}, the largest value of $T$ for which the map
has visible tori is given as a function of $\eta$. 
For these $T$ values, the central invariant curves are warped and twisted
(see \Fig{Figure9}a) and the surrounding two-tori have a small
poloidal diameter. For large $\eta$, the transition from integrability
to complete chaos is rapid.

\begin{table}
\caption{ Largest time, $T$, for which \Eq{eq:HexMap} has invariant tori
  as $\eta$ varies}
\begin{center}
\begin{tabular}{|c|c|c|c|c|c|c|c|c|c|c|}\hline
        $\eta$ & 0 & 5 & 15 & 25 & 35 & 45 & 55 & 65 & 75 & 85 \\ \hline
        $T$ & 1.81 & 1.8 & 1.7 & 1.6 & 1.5 & 1.35 & .97 & .6 & .6 & .25 \\ \hline
\end{tabular}\end{center}
\label{tbl:InvTori}
\end{table}

The evolution of the  toroidal invariant curve at the center of the
toroidal family of two-tori is shown in more detail in
\Fig{Figure9}. These curves are found using the center of mass
algorithm algorithm of \App{sec:algorithm}. In panel (a), the
curve is plotted for a number of different  parameter values.  As $T
\rightarrow 0$ the invariant curve approaches the closed curve
of the autonomous flow of \Eq{eq:HexVelField} given by
\begin{equation}\label{eq:Equilibria}
 \cos(y)+2\cos\left(\frac{\sqrt{3}x}{2}\right)\cos\left(\frac{y}{2}\right) = 0\;.
\end{equation}
This is shown as the darkest black curve at $z = 0$. For larger $T$ values the invariant curve
becomes increasingly distorted, and appears to lose
differentiability. The curve is destroyed for $T \approx 1.81$. For
larger values of $T$, as shown in panel (c), the phase portrait appears
to be completely chaotic. Panel (b) shows the longitudinal rotation number of the
invariant curves, $\omega$, as a function of $T$. As $T \rightarrow 0$,
$\omega \rightarrow 0$ since the mapping limits to a time $T$ map of the
autonomous flow. As $T$ increases the rotation number grows
monotonically. There are gaps in the rotation number corresponding to
parameter intervals where the curve is destroyed and reforms. 

The six poloidal tori shown in \Fig{Figure8} also appear to arise from
streamlines of the autonomous, hexagonal flow \Eq{eq:HexVelField} 
as $T$ increases from zero. Symmetry implies that they 
appear in the planes $x= 0$ and $y=0$ and their rotations
by $\pi/3$. When $\eta = 0$, the streamlines
of \Eq{eq:HexVelField} in these planes correspond to contours of
the functions
\begin{eqnarray}
 \Psi_1(0,y,z)&=&\cos(z)\sin^2\left(\frac{y}{2}\right)\left(\frac{1+2\cos(\frac{y}{2})}{1+\cos(\frac{y}{2})}\right)^2
 \label{eq:psi1}\\ 
 \Psi_2(x,0,z)&=&\cos(z)\sin^{\frac23}\left(\frac{\sqrt{3}x}{2}\right)\left(1-\cos\left(\frac{\sqrt{3}x}{2}\right)\right)^{\frac23} \;,
 \label{eq:psi2} 
\end{eqnarray}
respectively. When $\eta \neq 0$, a closed form 
expression for the streamlines is difficult to find. Using $T=10^{-4}$ and the center of
mass algorithm, computations show that the invariant curves at the
center of the six elliptic tori limit onto the streamline defined by $\Psi_1\approx 0.3868$
This streamline presumably is selected by a mechanism similar to the classical
Poincar\'e-Birkhoff island chain created in when an integrable area-preserving map is perturbed.  

Just as in the two-dimensional, island chain case, one of the $\Psi_2$ streamlines apparently
becomes a hyperbolic (saddle) invariant curve in the same manner that the
$\Psi_1 = 0.3868$ streamline becomes an elliptic invariant curve.
Indeed, numerical computations show localized chaos around the planes $y = 0$ and $y= \pm \sqrt{3}x$ (recall \Fig{Figure8}a and c) that grow as $T$ increases from zero.
Transverse intersections of the stable and unstable manifolds of these saddle
curves provide a mechanism for the onset of three-dimensional
chaos \cite{Lomeli03}.  

The disappearance of the elliptic invariant curves as $T$ increases is expected to be caused by various resonant bifurcations between their longitudinal and transverse rotation numbers. This hypothesis is supported by the observation that the regular region surrounding the elliptic curves shrinks as they approach destruction. We hope to report further on these bifurcations in a future paper.

\subsection{Transport}\label{sec:Levy}

While Lyapunov exponents measure the divergence of neighboring
trajectories, they give no information about how trajectories disperse
over macroscopic distances, that is about ``transport.'' The goal of this
section is to develop several measures of transport.

A first distinction to make is between trajectories that are bounded and
those that are not. One class of bounded orbits are those that remain
forever trapped in a single fundamental cell, $\sH$. These, for example,
include some of the invariant tori that we previously studied, though
it is possible for invariant tori to exist on scales larger than a
single cell. Since some trajectories never leave $\sH$, it is reasonable to restrict
the calculation to the set of initial conditions whose trajectories
eventually do leave; this set is equivalent, up to a set of zero volume,
to the subset that is ``accessible'' to trajectories that begin outside
$\sH$ \cite{Meiss1997}. The accessible set can be constructed by
considering the incoming set $\sH_{in}$ which is defined to be
\[
\sH_{in} = \sH \setminus F(\sH) \;,
\]
i.e., the portion of $\sH$ that does not intersect $F(\sH)$. According to
this definition, $F^{-1}(\sH_{in})$ is outside of $\sH$, thus every
point in $\sH_{in}$ has just entered $\sH$. The accessible portion of
$\sH$ is the portion that can be reached by starting in the incoming
set:
\[
\sH_{acc} = \bigcup\limits_{t=0}^{\infty} F^t (\sH_{in}) \cap \sH \;.
\]
In order to leave $\sH$, a trajectory must first land in the exit set
\[
\sH_{exit} = \sH \setminus F^{-1}(\sH) \;.
\]
By volume preservation, the incoming and exit sets have the same volume.

A simple measure of transport is the ``exit time distribution,''
$\psi(t)$, defined to be the probability that a trajectory that starts
in $\sH_{in}$ at $t=0$, first leaves $\sH$ at time $t$. To compute this
we cover $\sH$ with a grid of points. Each point is iterated with the
inverse map to see if it remains in $\sH$---if so, it is discarded. The
remaining points are in the incoming set, and they are iterated forward
until they exit $\sH$.

\begin{figure*}
\begin{center}
\includegraphics[width=6in]{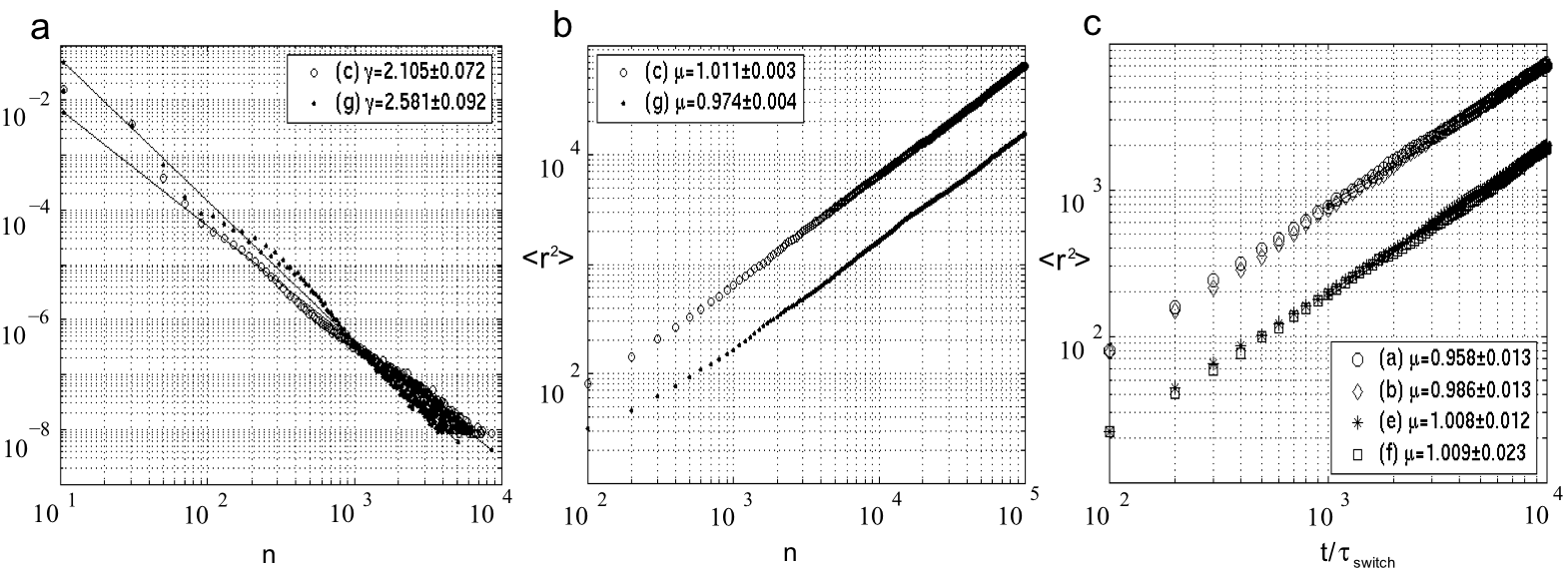}
\end{center}
\caption{Exit time and mean square displacement with parameters given in \Tbl{Table1}
  (a) Exit time distribution  for a $1000\times1000\times500$ grid in fundamental cell.
  (b) Mean-squared displacement for $104$ randomly sampled points in $\sH_{exit}$.
  (c) Mean-squared displacement for the flow \Eq{eq:VelField} with amplitudes
  given in for $103$ randomly sampled points in $\sH_{exit}$.}
\label{Figure10}
\end{figure*}

Exit time distributions for hyperbolic systems are typically
exponential; however, it is common in volume-preserving dynamics for $\psi$ to
have a power-law form \cite{Meiss1992}:
\begin{equation}\label{eq:ExitTimePdf}
\psi(t)\sim t^{-\gamma} \;.
\end{equation}
For the case of a volume-preserving map, it is a consequence of Kac's
theorem that  the average exit time
\begin{equation}\label{eq:exitTimePdf}
\tau_{exit}=\int_0^{\infty}t\psi(t)dt \;,
\end{equation}
exists \cite{Meiss1997}. Thus $\psi(t) = o(t^{-2})$ as $t \rightarrow
\infty$. The computations of exit time distributions, shown in
\Fig{Figure10}a, are consistent with \Eq{eq:ExitTimePdf} and
$\gamma \approx 2.1-2.5$.  In each case, as the theory requires, $\tau_{exit}<\infty$ since
the distribution has power law tail with $\gamma > 2$.

Chaotic trajectories that leave the fundamental cell tend to have
unbounded orbits (though as we have seen there can also be
superstructures extending over several copies of $\sH$). One measure of
the transport properties of unbounded trajectories is the growth rate of
the mean-squared displacement, i.e.
\begin{equation}\label{eq:avegrowth1}
        \ave{r^2(t)}\sim t^{\mu} \;.
\end{equation}
The case $\mu=1$ corresponds to diffusive motion; however, many
dynamical systems exhibit either subdiffusive ($\mu<1$) or
superdiffusive ($\mu>1$) behavior \cite{Klafter1980}.
Anomalous diffusion has been observed in many systems, for example, in two \cite{Solomon1994,Weeks1996}
and three-dimensional fluids \cite{Fogleman2001}, plasma physics \cite{del-Castillo2004},
and electron transport on semiconductor superlattices \cite{Fromhold2004}.

One phenomenological model that gives anomalous diffusion of the form of
\Eq{eq:avegrowth1} is a continuous time random walk.  In this model
particles  are assumed to move from one point to another under a
sequence of ``L\'evy flights'' whose length and duration are taken from
probability distributions with power law tails. Such models display both
normal and superdiffusive behavior
\cite{Klafter1987,Zumofen1990,Klafter1992}. A trapping
probability distribution, separating jumps by trapping events in which
the particle remains at its current position for a randomly selected
time can also be added to the model---in our case this distribution is
$\psi(t)$. For this situation, the dynamics can also exhibit subdiffusive
behavior. Subdiffusion occurs when the motion is dominated by trapping
events, and superdiffusion when flights are the more influential factor
\cite{Klafter1995}.

For our model the average exit time is finite; consequently, if there
are no flights, one can expect normally diffusive behavior \cite{Montroll1969}.
Though it is not easy to directly measure the distribution of
particles undergoing so-called ``L\'evy flights,''
we can compute the mean squared growth rate $\mu$, \Eq{eq:avegrowth1}.
To do this we randomly sample $10^4$ points in $\sH_{exit}$, iterating each
$10^5$ times.  We also compute the mean square displacement for the
continuous system, though in this case, since much
computation is involve we only use $10^3$ initial points and integrate
for $10^4$ periods. In both cases the results, shown in \Fig{Figure10}, are
consistent with $\mu = 1$ however $\mu=1$ does not always fall within
the $90\%$ confidence interval. However, the least squares technique
for computing the asymptotic growth rate may not give an accurate
estimate of the error. In general though, it appears that these models
are normally diffusive and that there are no L\'evy flights, unlike
many similar dynamical systems.

\section{Discussion}\label{sec:conclusions}

We have studied two idealized, kinematic models of the K\"uppers-Lortz
state. The building blocks of the models are vortical roll arrays
whose axes are rotated through an angle $\Theta$ with respect to one
another. In the continuous-time construction, the time dependence of
the amplitude of the rolls is controlled by a stochastic Busse-Heikes
system \Eq{eq:BH}. When the amplitudes of the rolls are step functions
and the rolls switch instantaneously from one set to another, we
derived an analytical map, \Eq{eq:HexMap}.

At the simplest level, the transition to chaos is governed
by the the ratio of switching time,  $\tau_\mathrm{switch}$, to
turnover turnover time, $\tau_\mathrm{roll}$. As this parameter goes
to zero, the flow approaches the autonomous, integrable case and when it is
sufficiently small, the dynamics is similar, showing the classical
hexagonal patterns observed in rotating Rayleigh-B\'enard convection.

As this ratio grows, both the flow and the map have a family of
invariant tori surrounding an elliptic invariant curve.
Outside these tori is a chaotic sea. The tori undergo a complex
sequence of bifurcations as the parameters change that we hope
to report on in a future paper. In some cases we have also observed localized
mixing for trajectories trapped within an invariant torus.  This
occurs, for example, if the central invariant curve is temporarily destroyed
by a bifurcation, but the surrounding two-tori persist. 

For large values of the the switching time,  the Lyapunov
phase portraits indicate nearly complete mixing. We characterized
the transport properties of our system by computing the mean exit-time 
from the fundamental cell and the mean-squared drift of trajectories over the
extended domain. Computations for the blinking-roll mapping and the
flow showed algebraic decay of the exit time distribution, and the mean-squared growth consistent with normal diffusion. These results, in conjunction with
anomalous diffusion theory, suggest that there is no analogue of the open streamlines
in an autonomous flow or the ``accelerator modes'' of area-preserving maps that would effect long range, superdiffusive transport.

We also showed that the dynamics of the flow and map models is highly
correlated even when the switching time is far from
instantaneous. This observation suggests that a much broader class of
continuous-time models, perhaps with periodic switching between more
than three states, can be accurately modeled with the blinking roll mapping.
With this in mind, one could easily extend the blinking roll system
to include more realistic spatial descriptions for the flow, like
Chandrasekhar functions, which are solutions to the linearized
Boussinesq approximation with a no-slip lower boundary condition.
However, it would be more difficult to obtain an explicit mapping
and the computations would consequently be slower. For the simple case
explored here, the Lyapunov phase portraits can be computed about twenty times
faster for the map than for the flow.

The mapping has tori over a wide range of switching and tilt angles.
We explored some of their behavior, but many mysteries remain. For
instance, the middle Lyapunov exponent appears to be zero to numerical
accuracy for both the continuous and discrete models. While this would
be rigorously true for autonomous flows or reversible maps, we do not
have an explanation of this for our systems. 
In addition, it would be interesting to study in detail the bifurcations that give
rise to the families of invariant tori and curves, as well as those that
ultimately lead to their destruction.  Numerical studies show that destruction of
invariant curves and tori typically enhances the mixing, but as of yet there is no
satisfactory theory of transport for these three-dimensional systems.

\appendix

\section{Appendix: Mean Roll Turnover Time}\label{sec:meanT}

Here, we compute the average roll turnover time, $\tau_{\text{roll}}$,
for a single roll with streamfunction $\psi=\sin(y)\cos(z)$. Suppose an
initial condition is placed along the line $z=0$,
$0<y=y_0<\frac{\pi}{2}$. The flow along this streamline is given by
\Eq{eq:TiltMap} (with $\eta=0$): 
\begin{equation} 
  \Phi_{t} \left(x, y_0, 0\right) = 
  \left(x,\sgn(y_0)\cos^{-1}\left(\frac{\cos(y_0)\cn(t)\dn(t)}{1-\cos^2(y_0)\sn^2(t)}\right),\sin^{-1}\left(-\sn(t)\cos(y_0)\right)\right) \;. 
\end{equation}
Now, solve for the $t$ value such that
\begin{equation*}
  \Phi_{t} \left(x, y_0, 0\right) = \left(x, \frac{\pi}{2}, y_0-\frac{\pi}{2}\right) \;.
\end{equation*}
This corresponds to one quarter of a rotation around a given
streamfunction. This is possible since the tracers rotate
counterclockwise in the domain $\left\{(x,y):0\leq y\leq \pi,
  -\frac{\pi}{2}\leq z\leq\frac{\pi}{2}\right\}$. From the third
component, we find a simple equation for $t$: 
\begin{equation*}
  t=\sn^{-1}(1,k)=F\left(\frac{\pi}{2},k\right)\;.
\end{equation*}
Here, the modulus of the elliptic functions is \Eq{eq:modulus} and
$F(\frac{\pi}{2},k)$ is the complete elliptic integral of 
the first kind. To get the average roll turnover time for all
streamfunctions, we integrate over $k$ to find \cite{Byrd1954}: 
\begin{equation}\label{eq:Troll}
  \tau_{\text{roll}}=4\int_{0}^{1}F\left(\frac{\pi}{2},k\right)dk=\pi^2 \nonumber \;.
\end{equation}
It is straightforward to show that $\tau_{roll}$ is independent of $\eta$.

\section{Appendix: Finding Invariant Curves}\label{sec:algorithm}

Locating invariant curves is a more challenging task than finding fixed
or periodic points of a map.  For a map, the dynamics on an
invariant curve is typically conjugate to a rigid rotation with
irrational winding number. Therefore,
one cannot use standard root finding techniques since the orbit never
returns to itself. Though there are many standard algorithms for finding
periodic orbits, methods for finding quasiperiodic ones are not as well developed. One algorithm uses Newton's method coupled
with an interpolation algorithm to find a single point on an invariant
curve \cite{Castella2000}. In another, the curve is approximated by a finite
Forier series and Newton's method is 
used to solve for the coefficients \cite{Jorba2001}. This 
this method is hindered by the assumption of conjugacy to rigid rotation number with an
a priori known rotation number.

We use a simple algorithm that computes a single point on the invariant
curve; however, only one point is needed since the rest of the invariant
curve can be found through iteration. The method is limited because it
only works for the case of an elliptic invariant curve, and it is
rather slow. On the other hand, the method has significant advantages in
that it does not require foreknowledge of the rotation number $\omega$
and it is easy to implement. 

The first step is to choose an appropriate plane that is expected to intersect 
the invariant curve transversely. We fatten this plane into a thin slice of width
$\epsilon$ and 
choose a point $\Bx_0$ in the slice close to where the
curve is expected. Iterate the initial
point and record the times it lands within the slice. Once $N$
such near returns are found, we compute the center of mass of these
points, $\bar{\Bx}$. If the elliptic curve is surrounded by
a family of two-tori, and these have a convex intersection with the
slice then $\bar{\Bx}$ will approximate the position of the enclosed
invariant curve. We now decrease $\epsilon$ and repeat the process with $\Bx_0=\bar{\Bx}$ as
the new starting position. This procedure appears to work well to find a point on the
invariant curve to machine precision. Given this point, we can now iterate to find the
longitudinal rotation number $\omega$ of the curve. 

\clearpage

\bibliographystyle{unsrt}

\end{document}